\documentclass[11pt]{article}
\font\grb=eurb10

\def\bphi{\hbox{\grb\char'047}\,}

\def\k{\hbox{\bf k}}

\def\u{\hbox{\bf u}}
\def\v{\hbox{\bf v}}

\begin{document}

\title{Infinite hierarchies of exact solutions of the Einstein and
Einstein--Maxwell equations for interacting waves and inhomogeneous
cosmologies}

\author{G. A. Alekseev$^{1,}$\thanks{Email address: {\tt G.A.Alekseev@mi.ras.ru}}
\ and J. B. Griffiths$^{2,}$\thanks{E--mail: {\tt J.B.Griffiths@Lboro.ac.uk}}
\\ \\ $^1$Steklov Mathematical Institute,
Gubkina 8, Moscow 117966, GSP-1, Moscow, Russia.
 \\ 
$^2$Department of Mathematical Sciences, Loughborough University, Loughborough, \\
Leics. LE11 3TU, U.K. \\ }

\date{\today}
\maketitle
\begin{abstract}
For space-times with two spacelike isometries, we present infinite hierarchies of
exact solutions of the Einstein and Einstein--Maxwell equations as represented by their
Ernst potentials. This hierarchy contains three arbitrary rational functions of an
auxiliary complex parameter. They are constructed using the so called ``monodromy
transform'' approach and our new method for the solution of the linear singular
integral equation form of the reduced Einstein equations. The solutions presented,
which describe inhomogeneous cosmological models or gravitational and electromagnetic
waves and their interactions, include a number of important known solutions as
particular cases.
\end{abstract}

\subsection*{Introduction}

A number of solution-generating techniques are known which provide tools
for the construction of vacuum and electrovacuum solutions of Einstein's
equations for space-times with symmetries.  These methods are based on the
integrability of the symmetry reduced Einstein equations (viz. the Ernst equations).
However, most of them were primarily designed to construct exact stationary
axisymmetric solutions for which an additional regularity condition should be
satisfied on the axis. This condition does not apply to interacting waves or
cosmological models as considered here.

Apart from the completely linearizable subcase of Einstein--Rosen vacuum
gravitational waves, the only techniques which provide nontrivial tools
for the construction of solutions for the dynamical case are the vacuum
Belinskii--Zakharov inverse-scattering method \cite{BelZak78}, the so
called ``monodromy transform'' approach \cite{Alek85,Alek87,Alek99}, and
the group-theoretical approach recently developed by Hauser and Ernst
\cite{HauErn99}. In particular, the methods of \cite{BelZak78} enable the
construction of soliton perturbations of homogeneous cosmological models
and some specific solutions for wave interaction regions. For example,
the Khan--Penrose \cite{KhaPen71} or Nutku--Halil \cite{NutHal77}
solutions for the interaction region for colliding impulsive
gravitational waves on a Minkowski background formally turn out to be
two-soliton solutions on a symmetric Kasner background.

Here we consider the monodromy transform approach and the linear singular
integral equations which arise in this context as an alternative form of
the reduced Einstein equations. We present a new method for the solution of
these equations which gives rise to infinite hierarchies of exact
solutions. Among many other solutions, these include the particular cases
mentioned above together with other soliton solutions on the symmetric
Kasner background and their non-soliton extensions.

\subsection*{Integral equation form of reduced Einstein equations}

According to methods developed in \cite{Alek85,Alek87,Alek99}, any
solution of the Ernst equations can be constructed from the solution of
the linear singular integral equation
 \begin{equation}
{1\over \pi i}\int_L
{[\lambda ]_\zeta\over\zeta-\tau}
{\cal H}(\tau,\zeta)
{\bphi}(\xi,\eta,\zeta)\, d\zeta =-\k(\tau)
\label{IntEqs}
\end{equation}
considered here for the hyperbolic case only.
The parameters $\xi$, $\eta$ are two real null space-time coordinates,
e.g. $(\xi,\eta)=(x+t,x-t)$.
These coordinates span some local region in the neighbourhood of some
initial regular
space-time point $P_0$: $\xi=\xi_0$, $\eta=\eta_0$, in which local
solutions of the reduced Einstein equations are considered.

The integration in (\ref{IntEqs}) is performed along the path $L$
on the spectral plane $w$ which consists of two disconnected parts
$L_{\scriptscriptstyle+}$ and $L_{\scriptscriptstyle-}$.
In the hyperbolic case, these are chosen as the segments of
the real axis in the $w$-plane, which go from $w=\xi_0$ to
$w=\xi$, and from $w=\eta_0$ to $w=\eta$ respectively.(We choose
$\xi_0\ne \eta_0$ and take $\xi$ and $\eta$ sufficiently close to
$\xi_0$ and $\eta_0$ that the segments $L_{\scriptscriptstyle\pm}$
do not overlap.)

The integral in (\ref{IntEqs}) splits into two, one of which possesses a
singular kernel of Cauchy type and should be understood as a Cauchy
principal value integral.  The integration parameter $\zeta$ and a
parameter $\tau$ span both of the contours $L_{\scriptscriptstyle+}$ and
$L_{\scriptscriptstyle-}$. Sometimes it will be convenient to introduce
suffices: $\zeta_{\scriptscriptstyle+},\tau_{\scriptscriptstyle+}\in
L_{\scriptscriptstyle+}$ and
$\zeta_{\scriptscriptstyle-},\tau_{\scriptscriptstyle-}\in
L_{\scriptscriptstyle-}$.

In the integrand in (\ref{IntEqs}), $[\lambda]_\zeta={1\over
2}(\lambda_{\rm left}-\lambda_{\rm right})$.  This represents the
jump on the contour, i.e.  half of the difference between left
and right limit values at the point $\zeta\in
L_{\scriptscriptstyle+}$ or $\zeta\in L_{\scriptscriptstyle-}$ of
some ``standard'' function $\lambda(\xi,\eta,w)$.  This function
is a product of two functions
$\lambda(\xi,\eta,w)=\lambda_{\scriptscriptstyle+}(\xi,w)
\lambda_{\scriptscriptstyle-}(\eta,w)$ given by
 \begin{equation}
\label{lampm}
\lambda_{\scriptscriptstyle+}=\sqrt{w-\xi\over w-\xi_0},
\qquad \lambda_{\scriptscriptstyle-}=\sqrt{w-\eta\over w-\eta_0},
\end{equation}
 with the additional conditions
$\lambda_{\scriptscriptstyle+}|_{w=\infty}
=\lambda_{\scriptscriptstyle-}|_{w=\infty}=1$.
Each of these functions is an analytic function on the whole
spectral plane $w$ apart from the cut $L_{\scriptscriptstyle+}$
or $L_{\scriptscriptstyle-}$ respectively, whose endpoints are the
branching points of the corresponding function.

In the equations (\ref{IntEqs}), the three-dimensional complex vector
function $\bphi(\xi,\eta,\zeta)$ is unknown, and the right hand
side $\k(\tau)$ is a three-dimensional
complex vector function of the spectral parameter which may be taken to be
 \begin{equation}
\k(w)=\{1,{\bf u}(w),\v(w)\},
\label{kw}
\end{equation}
 where ${\bf u}(w)$ and $\v(w)$ are arbitrary functions.
 The kernel of the integral in (\ref{IntEqs}) is a scalar
function ${\cal H}(\tau,\zeta)$ given by
\begin{eqnarray}
 &&{\cal H}(\tau,\zeta) =1+i (\zeta-\beta_0) ({\bf
 u}(\tau)-{\bf u}^\dagger(\zeta)) +\alpha_0^2{\bf u}(\tau){\bf
u}^\dagger(\zeta) \nonumber \\
&&\qquad\qquad\qquad -4 (\zeta-\xi_0)(\zeta-\eta_0)\v(\tau)\v^\dagger(\zeta)
\label{calH} \end{eqnarray}
where the dagger denotes complex conjugation: e.g.
${\bf u}^\dagger(w)\equiv \overline{{\bf u}(\overline{w})}$. The additional
constants in (\ref{calH}) are $\alpha_0=(\xi_0-\eta_0)/2$ and
$\beta_0=(\xi_0+\eta_0)/2$.

It is important to emphasize that the integral equations (\ref{IntEqs}), and
hence the functions ${\bf u}(w)$, $\v(w)$ and $\bphi(\xi,\eta,w)$, only need
to be evaluated on the two cuts $L_{\scriptscriptstyle+}$ and
$L_{\scriptscriptstyle-}$ in the spectral plane.  Thus all the above vector
and scalar functions of the spectral parameter are actually determined by
pairs of functions which represent their values on these contours.  For
convenience we shall denote the values of these functions on
$L_{\scriptscriptstyle\pm}$ by the corresponding suffices:
 \begin{equation}
\{\u(w),\v(w)\}=
\left\{\matrix{\{\u_{\scriptscriptstyle+}(w),
\v_{\scriptscriptstyle+}(w)\},\qquad w\in
L_{\scriptscriptstyle+}\cr
\{\u_{\scriptscriptstyle-}(w),
\v_{\scriptscriptstyle-}(w)\},\qquad w\in
L_{\scriptscriptstyle-}}\right.
 \label{defuvpm}
\end{equation}
Thus, in (\ref{IntEqs}) written in a more explicit form, we
actually have two unknown vector functions
$\bphi_{\scriptscriptstyle\pm}$. For any of these suffixed
functions we can use also an alternative definition, for example,
 $$ \bphi(\xi,\eta,\tau_{\scriptscriptstyle\pm})\equiv
\bphi_{\scriptscriptstyle\pm}(\xi,\eta,\tau). $$

Using this notation, it is convenient to split the
integral in (\ref{IntEqs}) into separate integrals over
$L_{\scriptscriptstyle+}$ and $L_{\scriptscriptstyle-}$ and
to consider separately the cases
$\tau=\tau_{\scriptscriptstyle+}\in
L_{\scriptscriptstyle+}$ and $\tau=\tau_{\scriptscriptstyle-}\in
L_{\scriptscriptstyle-}$. It is also convenient to denote the four
scalar kernels which appear in the integrands of (\ref{IntEqs})
in the form
 \begin{eqnarray}
{\cal H}(\tau_{\scriptscriptstyle+},\zeta_{\scriptscriptstyle+})\equiv
{\cal H}_{\scriptscriptstyle++}(\tau,\zeta), && \qquad {\cal
H}(\tau_{\scriptscriptstyle+},\zeta_{\scriptscriptstyle-})\equiv
{\cal H}_{\scriptscriptstyle+-}(\tau,\zeta), \nonumber \\
 {\cal H}(\tau_{\scriptscriptstyle-},\zeta_{\scriptscriptstyle+})\equiv
{\cal H}_{\scriptscriptstyle-+}(\tau,\zeta), && \qquad {\cal
H}(\tau_{\scriptscriptstyle-},\zeta_{\scriptscriptstyle-})\equiv
{\cal H}_{\scriptscriptstyle--}(\tau,\zeta)
 \nonumber
 \end{eqnarray}
 where the functions ${\cal H}_{\scriptscriptstyle++}(\tau,\zeta)$,
${\cal H}_{\scriptscriptstyle+-}(\tau,\zeta)$,
${\cal H}_{\scriptscriptstyle-+}(\tau,\zeta)$ and
${\cal H}_{\scriptscriptstyle--}(\tau,\zeta)$ can be determined
explicitly in terms of the four functions
${\bf u}_{\scriptscriptstyle\pm}(w)$ and
$\v_{\scriptscriptstyle\pm}(w)$ using (\ref{calH}).

To conclude our description of the structure of the master
integral equations, we recall that the four functions
${\bf u}_{\scriptscriptstyle\pm}(w)$ and
$\v_{\scriptscriptstyle\pm}(w)$ appearing in (\ref{kw}) and (\ref{defuvpm})
play a significant role in the entire construction.  They determine
completely the
coefficients of the integral equations in the electrovacuum case.  In the
vacuum case there are only two such functions
${\bf u}_{\scriptscriptstyle\pm}(w)$, as
$\v_{\scriptscriptstyle\pm}(w)\equiv0$.  As shown in \cite{Alek87}, they
characterize unambiguously every individual solution of the Ernst equations.
Moreover, the singular integral equations (\ref{IntEqs}) possess a
unique solution for any given choice of analytical functions
${\bf u}_{\scriptscriptstyle\pm}(w)$ and
$\v_{\scriptscriptstyle\pm}(w)$.

We recall now also, that the general local solution of the 
hyperbolic Ernst equations can be expressed by quadratures in terms 
of the solution of (\ref{IntEqs}) 
\begin{eqnarray} &&{\cal E} =-1 
-{2\over\pi} \int_L [\lambda]_\zeta \big[1-i (\zeta-\beta_0) {\bf 
u}^\dagger(\zeta)\big] \bphi^{[u]}(\xi,\eta,\zeta)\,d\zeta \nonumber 
\\ &&\Phi ={2\over\pi} \int_L [\lambda]_\zeta \big[1-i 
(\zeta-\beta_0) {\bf u}^\dagger(\zeta)\big] 
\bphi^{[v]}(\xi,\eta,\zeta)\,d\zeta
\label{Potentials}
\end{eqnarray}
where $\bphi^{[u]}$ and $\bphi^{[v]}$, in some association with
the definition (\ref{kw}), denote respectively the second and
third components of the vector solutions
$\bphi$ of the master integral equation
(\ref{IntEqs}), corresponding to a given choice of the monodromy
data functions ${\bf u}_{\scriptscriptstyle\pm}(w)$ and
$\v_{\scriptscriptstyle\pm}(w)$.
In a more explicit form, each of the
the integrals in (\ref{Potentials}) should be split into two
integrals evaluated over $L_{\scriptscriptstyle+}$ and
$L_{\scriptscriptstyle-}$.

\subsection*{New hierarchies of solutions }

Here we will construct a class of hyperbolic solutions 
that is determined by the rational monodromy data 
\begin{equation}\label{uvpm}
\u_{\scriptscriptstyle\pm}(w) ={U_{\scriptscriptstyle\pm}(w)\over
Q_{\scriptscriptstyle\pm}(w)},\qquad
\v_{\scriptscriptstyle\pm}(w) ={V_{\scriptscriptstyle\pm}(w)\over
Q_{\scriptscriptstyle\pm}(w)}
\label{rational}
\end{equation}
 where $U_{\scriptscriptstyle+}(w)$,
$V_{\scriptscriptstyle+}(w)$,
$Q_{\scriptscriptstyle+}(w)$ and $U_{\scriptscriptstyle-}(w)$,
$V_{\scriptscriptstyle-}(w)$, $Q_{\scriptscriptstyle-}(w)$ are
arbitrary complex polynomials, provided $\u_+(w)$, $\v_+(w)$ and
$\u_-(w)$, $\v_-(w)$ do not have poles on $L_+$ and $L_-$ respectively.

For what follows, it is convenient to calculate some auxiliary polynomials
of two variables -- we introduce the four polynomials
$P_{\scriptscriptstyle\pm\pm}(\tau,\zeta)$ defined by the relations
\begin{equation}
{\cal H}_{\scriptscriptstyle\pm\dot\pm}(\tau,\zeta)
 ={P_{\scriptscriptstyle\pm\dot\pm}(\tau,\zeta)\over
Q_\pm(\tau) Q_{\dot\pm}^\dagger(\zeta)},
\label{Hpmpm}
 \end{equation}
 and four polynomials $R_{\scriptscriptstyle\pm\pm}(\tau,\zeta)$
defined from
them by
 \begin{equation}
 R_{\scriptscriptstyle\pm\dot\pm}(\tau,\zeta)=
{P_{\scriptscriptstyle\pm\dot\pm}(\tau,\zeta)-
P_{\scriptscriptstyle\pm\dot\pm}(\zeta,\zeta)\over \zeta-\tau}.
\label{Rpmpm}
\end{equation}
 In these definitions there are two sets of suffices, denoted as
dotted and
undotted,
which should each be taken to be the same. \ Finally, it is
convenient to
introduce
a redefinition of the unknown functions\begin{eqnarray}
&&\bphi_{\scriptscriptstyle+}(\zeta)=
-{ \lambda_{\scriptscriptstyle-}^{-1}(\zeta)
Q_{\scriptscriptstyle+}^\dagger(\zeta)
\over P_{\scriptscriptstyle++}(\zeta,\zeta)}
\widetilde{\bphi}_{\scriptscriptstyle+}(\zeta), \nonumber \\
&&\bphi_{\scriptscriptstyle-}(\zeta)=
-{ \lambda_{\scriptscriptstyle+}^{-1}(\zeta)
Q_{\scriptscriptstyle-}^\dagger(\zeta)
\over P_{\scriptscriptstyle--}(\zeta,\zeta)}
\widetilde{\bphi}_{\scriptscriptstyle-}(\zeta).
\label{Phitilde}
\end{eqnarray}
Hereafter we do not show explicitly the arguments $\xi$ and $\eta$ of
     $\bphi_\pm$ and $\lambda$ or the suffices $\pm$ at the points $\zeta$
     and $\tau$, unless it is necessary.

A direct substitution of (\ref{uvpm}) into equations
(\ref{IntEqs}) with the use of (\ref{Hpmpm})--(\ref{Phitilde}) leads
to the following convenient form of linear equations with polynomial right
hand sides
 \begin{eqnarray}\label{PolyEqs}
{1\over \pi i} \int\limits_{\xi_0}^{\xi}
{[\lambda_{\scriptscriptstyle+}]_\zeta
\over\zeta-\tau_{\scriptscriptstyle+}}
\widetilde{\bphi}_{\scriptscriptstyle+}(\zeta)\,d\zeta
&&=-{1\over \pi i} \int\limits_{\xi_0}^{\xi}
[\lambda_{\scriptscriptstyle+}]_\zeta
{R_{\scriptscriptstyle++}(\tau_{\scriptscriptstyle+},\zeta)\over
P_{\scriptscriptstyle++}(\zeta,\zeta)}
\widetilde{\bphi}_{\scriptscriptstyle+}(\zeta)\,d\zeta\nonumber
\\
&&\qquad-{1\over \pi i}
 \int\limits_{\eta_0}^{\eta}
[\lambda_{\scriptscriptstyle-}]_\zeta
{R_{\scriptscriptstyle+-}(\tau_{\scriptscriptstyle+},\zeta)
\over P_{\scriptscriptstyle--}(\zeta,\zeta)}
\widetilde{\bphi}_{\scriptscriptstyle-}(\zeta)\,d\zeta
 +\pmatrix{Q_{\scriptscriptstyle+}(\tau_{\scriptscriptstyle+})\cr
U_{\scriptscriptstyle+}(\tau_{\scriptscriptstyle+})\cr
V_{\scriptscriptstyle+}(\tau_{\scriptscriptstyle+})}, \nonumber
\\ &&{} \label{lineqs} \\
{1\over \pi i}
\int\limits_{\eta_0}^{\eta}{[\lambda_{\scriptscriptstyle-}]_\zeta
\over\zeta-\tau_{\scriptscriptstyle-}}
\widetilde{\bphi}_{\scriptscriptstyle-}(\zeta)\,d\zeta
&&=-{1\over \pi i} \int\limits_{\eta_0}^{\eta}
[\lambda_{\scriptscriptstyle-}]_\zeta
{R_{\scriptscriptstyle--}(\tau_{\scriptscriptstyle-},\zeta)\over
P_{\scriptscriptstyle--}(\zeta,\zeta)}
\widetilde{\bphi}_{\scriptscriptstyle-}(\zeta)\,d\zeta\nonumber
\\
&&\qquad-{1\over \pi i}
 \int\limits_{\xi_0}^{\xi} [\lambda_{\scriptscriptstyle+}]_\zeta
{R_{\scriptscriptstyle-+}
(\tau_{\scriptscriptstyle-},\zeta)\over
P_{\scriptscriptstyle++}(\zeta,\zeta)}
\widetilde{\bphi}_{\scriptscriptstyle+}(\zeta)\,d\zeta
+\pmatrix{Q_{\scriptscriptstyle-}(\tau_{\scriptscriptstyle-})\cr
U_{\scriptscriptstyle-}(\tau_{\scriptscriptstyle-})\cr
V_{\scriptscriptstyle-}(\tau_{\scriptscriptstyle-})} \nonumber
  \end{eqnarray}
 if we impose constraints on the coefficients of
the rational functions (\ref{uvpm}) such that
 \begin{equation}
P_{\scriptscriptstyle+-}(\zeta,\zeta)
=P_{\scriptscriptstyle-+}(\zeta,\zeta)=0.
\label{ansatz1}
 \end{equation}
 This leads to a large class of explicit solutions
$\widetilde{\bphi}_{\scriptscriptstyle\pm}(\xi,\eta,\tau)$
of (\ref{lineqs}) that are regular on the cuts $L_\pm$. However, the
solution of the Ernst equations needs the solutions
$\bphi_{\scriptscriptstyle\pm}(\xi,\eta,\tau)$ of (\ref{IntEqs}) to be
regular on the cuts $L_\pm$. Fortunately, all additional singularities
(poles) of $\bphi_{\scriptscriptstyle+}(\xi,\eta,\tau)$ on $L_+$ and
$\bphi_{\scriptscriptstyle-}(\xi,\eta,\tau)$ on $L_-$, which
arise from the denominators in (\ref{Phitilde}), can be avoided by the
additional restrictions that $\u_+(\eta_0)=-i/\alpha_0$ and
$\u_-(\xi_0)=i/\alpha_0$. We therefore specify
 \begin{eqnarray}
&&\u_+(w)=-{i\over\alpha_0}+(w-\eta_0) {C_+(w)\over Q_+(w)} \nonumber\\
&&\u_-(w)={i\over\alpha_0}+(w-\xi_0) {C_-(w)\over Q_-(w)}
 \label{upm}
 \end{eqnarray}
 where $C_+(w)$, $C_-(w)$, $Q_+(w)$ and $Q_-(w)$ are arbitrary polynomials.
With these, the ansatz (\ref{ansatz1}) leads to the constraint
$C_-(w)=B(w)C_+^\dagger(w)-4 i A(w)V_+^\dagger(w)$ and, for the polynomials
in (\ref{rational}), the general solution of (\ref{ansatz1}) reads
 \begin{eqnarray}
&&U_+(w)= -\displaystyle{i\over\alpha_0}Q_+(w)+(w-\eta_0) C_+(w) \nonumber\\
&&U_-(w)= B(w)\left(\displaystyle{i\over\alpha_0}Q_+^\dagger(w)+(w-\beta_0)
C_+^\dagger(w)\right) 
 -4 i (w-\xi_0) A(w) V_+^\dagger(w) \nonumber\\
&&V_-(w)=A(w)\left(Q_+^\dagger(w)-i\alpha_0^2 C_+^\dagger(w)\right)
\nonumber\\
&&Q_-(w)=B(w)\left(Q_+^\dagger(w)-i\alpha_0^2 C_+^\dagger(w)\right)
 \label{evacpoly}
 \end{eqnarray}
 where the polynomials $A(w)$, $B(w)$, $C_+(w)$, $V_+(w)$ and $Q_+(w)$ can
be chosen arbitrarily, provided the corresponding functions $\u_\pm(w)$,
$\v_\pm(w)$ have no poles on the cuts $L_+$ and $L_-$ respectively.
The vacuum case, which occurs when $A(w)=V_+(w)=0$ and $B(w)=1$, yields
simpler expressions which involve just two arbitrary polynomials $C_+(w)$
and $Q_+(w)$.

Returning to (\ref{PolyEqs}), we note that the integral operators in the
left hand sides can be inverted using the Poincar\'e--Bertrand formula
\cite{Gakhov} for singular integrals
 \begin{equation}
{1\over \pi i}
\int_L {[\lambda]_\zeta\over\zeta-\tau}
{\bphi}(\zeta)\,d\zeta = f(\tau) \quad\Leftrightarrow\quad
\bphi(\tau)={1\over \pi i}
\int_L {[\lambda^{-1}]_\zeta\over\zeta-\tau} f(\zeta)\,d\zeta.
 \label{PB}
 \end{equation}
 This can be applied to the integrals over
$L_{\scriptscriptstyle+}$ (using $\lambda_{\scriptscriptstyle+}$),
or over $L_{\scriptscriptstyle-}$ (using $\lambda_{\scriptscriptstyle-}$).

Since the right hand sides of (\ref{PolyEqs}) are polynomials in
$\tau$, the inversion (\ref{PB}) leads to the solution in the form
 \begin{equation}\label{sums}
\widetilde{\bphi}_{\scriptscriptstyle\pm}(\tau)
=\sum\limits_{k=0}^{N_{\scriptscriptstyle\pm}}
\pmatrix {\widetilde{q}_{k{\scriptscriptstyle\pm}}\cr
\widetilde{u}_{k{\scriptscriptstyle\pm}}\cr
\widetilde{v}_{k{\scriptscriptstyle\pm}}}
Z_{k{\scriptscriptstyle\pm}}(\tau)
 \end{equation}
 where $N_{\scriptscriptstyle+}$ and
$N_{\scriptscriptstyle-}$ are
the maxima
of the degrees of the polynomials $U_{\scriptscriptstyle+}$,
$V_{\scriptscriptstyle+}$, $Q_{\scriptscriptstyle+}$ and
$U_{\scriptscriptstyle-}$, $V_{\scriptscriptstyle-}$,
$Q_{\scriptscriptstyle-}$
respectively, $\widetilde{u}_{k{\scriptscriptstyle\pm}}$,
$\widetilde{v}_{k{\scriptscriptstyle\pm}}$,
$\widetilde{q}_{k{\scriptscriptstyle\pm}}$ are unknown $\tau$-independent
functions of $\xi$ and $\eta$, and $Z_{k{\scriptscriptstyle\pm}}(\tau)$ are
``standard'' functions given by
 \begin{equation}
 Z_{k{\scriptscriptstyle\pm}}(\tau)=\displaystyle{1\over \pi i}
\int_{L_\pm}{[\lambda_{\scriptscriptstyle\pm}^{-1}]_\zeta\over\zeta-
\tau}\,
\zeta^k\,d\zeta.
 \end{equation}
 All these functions (integrals) can be evaluated as the residues of their
integrands at $\zeta=\infty$ are polynomials in $\tau$ of degree $k$.

We note now, that the vector integral equations (\ref{IntEqs})
decouple into three pairs of equations -- one pair for each of
the three components of $\widetilde{\bphi}_{\scriptscriptstyle+}$
and the corresponding component of
$\widetilde{\bphi}_{\scriptscriptstyle-}$.  All these pairs of
equations possess the same kernels but different right hand
sides.  Therefore, substituting the expressions (\ref{sums})
into (\ref{PolyEqs}) and using (\ref{Rpmpm}) with (\ref{ansatz1}),
we get three decoupled algebraic systems, each of order
$(N+2)\times (N+2)$ where
$N=N_{\scriptscriptstyle+}+N_{\scriptscriptstyle-}$ and
for the sets of unknowns $\widetilde{q}_{k{\scriptscriptstyle\pm}}$,
$\widetilde{u}_{k{\scriptscriptstyle\pm}}$,
$\widetilde{v}_{k{\scriptscriptstyle\pm}}$ respectively.  However, in view
of (\ref{Potentials}), we need the solutions of two of these systems only:
\begin{equation}
\matrix{
\sum\limits_{{\scriptscriptstyle B}=0}^{N+1}
{\cal D}_{\scriptscriptstyle A B}
\widetilde{u}_{\scriptscriptstyle B}=u_{\scriptscriptstyle A},\cr
\sum\limits_{{\scriptscriptstyle B}=0}^{N+1}
{\cal D}_{\scriptscriptstyle A B}
\widetilde{v}_{\scriptscriptstyle B}
=v_{\scriptscriptstyle A}}\qquad
{\cal D}=\pmatrix {D_{\scriptscriptstyle++} &
D_{\scriptscriptstyle+-}\cr
D_{\scriptscriptstyle-+} & D_{\scriptscriptstyle--}}
\label{AlgSys}\end{equation}
 where the indices $A,B=0,1,\ldots N+1$. The column vectors
$u_{\scriptscriptstyle A}$,
$v_{\scriptscriptstyle A}$ (shown below as rows) are composed
of the coefficients of the
polynomials $U_{\scriptscriptstyle\pm }(\zeta)$ and
$V_{\scriptscriptstyle\pm}(\zeta)$:
 \begin{eqnarray}
&&u_{\scriptscriptstyle A}
=\{u_{0{\scriptscriptstyle+}},u_{1{\scriptscriptstyle+}},
\ldots,u_{N_{\scriptscriptstyle+}},u_{0{\scriptscriptstyle-}},
u_{1{\scriptscriptstyle-}},\ldots,u_{N_{\scriptscriptstyle-}}
\} \nonumber \\
&&v_{\scriptscriptstyle
A}=\{v_{0{\scriptscriptstyle+}},v_{1{\scriptscriptstyle+}},
\ldots,v_{N_{\scriptscriptstyle+}},v_{0{\scriptscriptstyle-}},
v_{1{\scriptscriptstyle-}},\ldots,v_{N_{\scriptscriptstyle-}}
\}. 
 \end{eqnarray}
 Similarly, we combine the coefficients
$\widetilde{u}_{k{\scriptscriptstyle\pm}}$,
$\widetilde{v}_{k{\scriptscriptstyle\pm}}$ in (\ref{sums})
to form the column vectors (rows)
 \begin{eqnarray}
&&\widetilde{u}_{\scriptscriptstyle A}(\xi,\eta)
=\{\widetilde{u}_{0{\scriptscriptstyle+}},
\widetilde{u}_{1{\scriptscriptstyle+}},
\ldots,\widetilde{u}_{N_{\scriptscriptstyle+}},
\widetilde{u}_{0{\scriptscriptstyle-}},
\widetilde{u}_{1{\scriptscriptstyle-}},\ldots,
\widetilde{u}_{N_{\scriptscriptstyle-}}\} \nonumber\\
&&\widetilde{v}_{\scriptscriptstyle A}(\xi,\eta)
=\{\widetilde{v}_{0{\scriptscriptstyle+}},\widetilde{v}_{1{
\scriptscriptstyle+}},
\ldots,\widetilde{v}_{N_{\scriptscriptstyle+}},
\widetilde{v}_{0{\scriptscriptstyle-}},\widetilde{v}_{1{
\scriptscriptstyle-}},
\ldots,\widetilde{v}_{N_{\scriptscriptstyle-}}\}
\end{eqnarray}

The matrix $\Vert{\cal D}\Vert$ consists of the blocks
$D_{\scriptscriptstyle++}$, $D_{\scriptscriptstyle+-}$,
$D_{\scriptscriptstyle-+}$, $D_{\scriptscriptstyle--}$ of
orders $(N_{\scriptscriptstyle+}+1)\times
(N_{\scriptscriptstyle+}+1)$,
$(N_{\scriptscriptstyle+}+1)\times
(N_{\scriptscriptstyle-}+1)$,
$(N_{\scriptscriptstyle-}+1)\times
(N_{\scriptscriptstyle+}+1)$ and
$(N_{\scriptscriptstyle-}+1)\times
(N_{\scriptscriptstyle-}+1)$ respectively.  
Their components are determined by the integrals:
\begin{eqnarray}
&& (D_{\scriptscriptstyle++})_{kl}(\xi)=\delta_{kl}+{1\over \pi i}
\int\limits_{\xi_0}^{\xi} [\lambda_{\scriptscriptstyle+}]_\zeta
{(R_{\scriptscriptstyle++})_k(\zeta)\over
P_{\scriptscriptstyle++}(\zeta,\zeta)}
Z_{l{\scriptscriptstyle+}}(\zeta)
\,d\zeta\nonumber\\
&&(D_{\scriptscriptstyle+-})_{kl}(\eta)={1\over \pi
i}\int\limits_{\eta_0}^{\eta}
[\lambda_{\scriptscriptstyle-}]_\zeta
{(R_{\scriptscriptstyle+-})_k(\zeta)\over
P_{\scriptscriptstyle--}(\zeta,\zeta)}
Z_{l{\scriptscriptstyle-}}(\zeta)
\,d\zeta\nonumber\\
&&(D_{\scriptscriptstyle-+})_{kl}(\xi)={1\over \pi
i}\int\limits_{\xi_0}^{\xi}
[\lambda_{\scriptscriptstyle+}]_\zeta
{(R_{\scriptscriptstyle-+})_k(\zeta)\over
P_{\scriptscriptstyle++}(\zeta,\zeta)}
 Z_{l{\scriptscriptstyle+}}(\zeta)
\,d\zeta \\
&&(D_{\scriptscriptstyle--})_{kl}(\eta)=\delta_{kl}+{1\over \pi i}
\int\limits_{\eta_0}^{\eta} [\lambda_{\scriptscriptstyle-}]_\zeta
{(R_{\scriptscriptstyle--})_k(\zeta)\over
P_{\scriptscriptstyle--}(\zeta,\zeta)}
Z_{l{\scriptscriptstyle-}}(\zeta)
\,d\zeta\nonumber
 \end{eqnarray}
 where $(R_{\scriptscriptstyle\pm\pm})_k$ are the coefficients in the
expansions
$R_{\scriptscriptstyle+\pm}(\tau,\zeta)
=\sum_{k=0}^{N_{\scriptscriptstyle+}}
(R_{\scriptscriptstyle+\pm})_k(\zeta) \tau^k$ and
$R_{\scriptscriptstyle-\pm}(\tau,\zeta)
=\sum_{k=0}^{N_{\scriptscriptstyle-}}
(R_{\scriptscriptstyle-\pm})_k(\zeta)\tau^k$.

To calculate the final expressions for the Ernst potentials, we need
to evaluate the additional sets of integrals
\begin{eqnarray}
J_{k{\scriptscriptstyle+}}(\xi) ={1\over\pi i} \int\limits_{\xi_0}^{\xi}
[\lambda_{\scriptscriptstyle+}]_\zeta
{Q_{\scriptscriptstyle+}^\dagger(\zeta)-i(\zeta-\beta_0)
U_{\scriptscriptstyle+}^\dagger (\zeta)\over
P_{\scriptscriptstyle++}(\zeta,\zeta)}
Z_{k{\scriptscriptstyle+}}(\zeta)
\,d\zeta\nonumber\\
J_{k{\scriptscriptstyle-}}(\eta) 
={1\over\pi i}\int\limits_{\eta_0}^{\eta}
[\lambda_{\scriptscriptstyle-}]_\zeta
{Q_{\scriptscriptstyle-}^\dagger(\zeta)
-i(\zeta-\beta_0)
U_{\scriptscriptstyle-}^\dagger (\zeta)\over
P_{\scriptscriptstyle--}(\zeta,\zeta)}
Z_{k{\scriptscriptstyle-}}(\zeta) \,d\zeta, \nonumber
\end{eqnarray}
and to combine them into one row vector
 \begin{equation}
 J_{\scriptscriptstyle A}
=\{J_{0{\scriptscriptstyle+}},J_{1{\scriptscriptstyle+}},
\ldots,J_{N_{\scriptscriptstyle+}},J_{0{\scriptscriptstyle-}},
J_{1{\scriptscriptstyle-}},\ldots,
J_{N_{\scriptscriptstyle-}}\}. 
 \end{equation}
 Let us also define two additional $(N+2)\times (N+2)$ matrices 
 \begin{equation}
 {\cal G}_{\scriptscriptstyle AB} ={\cal D}_{\scriptscriptstyle AB}
-2iu_{\scriptscriptstyle A} J_{\scriptscriptstyle B},\qquad
{\cal F}_{\scriptscriptstyle AB} ={\cal D}_{\scriptscriptstyle AB}-2 i
v_{\scriptscriptstyle A} J_{\scriptscriptstyle B}. 
 \end{equation}
 All integrals determining the components of the matrices ${\cal
G}_{\scriptscriptstyle AB}$, ${\cal F}_{\scriptscriptstyle AB}$ and
${\cal D}_{\scriptscriptstyle AB}$ can be evaluated in terms of
the residues of their integrands  at the zeros of
$P_{\scriptscriptstyle++}(w,w)$ and $P_{\scriptscriptstyle--}(w,w)$
and at $w=\infty$. We then have 
 \begin{equation}
 {\cal E}= -{\det\Vert{\cal G}_{\scriptscriptstyle AB}
\Vert \over \det\Vert{\cal D}_{\scriptscriptstyle AB}\Vert},
\qquad
\Phi ={\det\Vert{\cal F}_{\scriptscriptstyle AB}
\Vert \over \det\Vert{\cal D}_{\scriptscriptstyle AB}\Vert}, 
 \label{pots}
 \end{equation}
 which are the final expressions for the Ernst potentials. 
These solutions generally possess essentially nonlinear properties. They
are not trivial time-dependent analogues of any stationary axisymmetric
solutions with regular axis of symmetry which have different structures
of monodromy data. The expressions (\ref{pots}) generally are not rational
functions of $\xi$, $\eta$.

When evaluating explicit examples, it may be noted that solutions with a
diagonal metric occur when
${\bf u}_{\scriptscriptstyle\pm}^\dagger =-{\bf u}_{\scriptscriptstyle\pm}$.
The plane symmetric (type~D) Kasner metric with ${\cal E}=-\alpha/\alpha_0$
is obtained using the constants
${\bf u}_{\scriptscriptstyle+}=-i/\alpha_0$,
${\bf u}_{\scriptscriptstyle-}=i/\alpha_0$ and
${\bf v}_{\scriptscriptstyle\pm}=0$.
The Khan--Penrose solution \cite{KhaPen71} for colliding plane impulsive
gravitational waves is obtained with $\v_+(w)=\v_-(w)=0$ and 
 \begin{equation}
 {\bf u}_+(w) =i k_+{w-a_+\over w-b_+},\qquad
{\bf u}_-(w) =ik_-{w-a_-\over w-b_-} 
 \end{equation}
 when the constants $a_\pm$, $b_\pm$ and $k_\pm$ are real. The
nondiagonal Nutku--Halil solution \cite{NutHal77} for non-colinear
impulsive waves is obtained from the same expression using complex
constants. This explicitly demonstrates that the above method is
applicable to both the linear and nonlinear cases.

\bigskip
This work was partly supported by the EPSRC and by the
grants 99-01-01150 and 99-02-18415 from the RFBR.

\end{document}